\documentclass[a4paper]{article}
\usepackage[numbers]{natbib}

\usepackage{INTERSPEECH2021}

\usepackage{todonotes}
\usepackage{url}
\usepackage{float}
\usepackage[caption = false]{subfig}
\usepackage{adjustbox}

\newcommand{\embeddingFn}{\mathcal{F_\theta}}
\newcommand{\trainSet}{\mathcal{D}}
\newcommand{\vocab}{\mathcal{V}}
\newcommand{\librikws}{\text{LibriKWS}}
\newcommand{\sysname}{\text{KeySEM}}
\newcommand{\linetalmp}{\text{MTLEmb}}

\title{Teaching keyword spotters to spot new keywords with limited examples}

\name{Abhijeet Awasthi$^{1,2}$, Kevin Kilgour$^1$, Hassan Rom$^1$}
\address{
  $^1$Google Research, Switzerland \\
  $^2$Indian Institute of Technology Bombay, India
  }
\email{
awasthi@cse.iitb.ac.in,
\{kkilgour,hassanrom\}@google.com}
\begin{document}
\maketitle
\begin{abstract}
  Learning to recognize new keywords with just a few examples is essential for personalizing keyword spotting (KWS) models to a user's choice of keywords. However, modern KWS models are typically trained on large datasets and restricted to a small vocabulary of keywords, limiting their transferability to a broad range of unseen keywords.
  Towards easily customizable KWS models, we present KeySEM ({\bf Key}word {\bf S}peech {\bf EM}bedding), a speech embedding model pre-trained on the task of recognizing a large number of keywords. Speech representations offered by KeySEM are highly effective for learning new keywords from a limited number of examples. Comparisons with a diverse range of related work across several datasets show that our method achieves consistently superior performance with fewer training examples. Although KeySEM was pre-trained only on English utterances, the performance gains also extend to datasets from four other languages indicating that KeySEM learns useful representations well aligned with the task of keyword spotting. Finally, we demonstrate KeySEM's ability to learn new keywords sequentially without requiring to re-train on previously learned keywords. Our experimental observations suggest that KeySEM is well suited to on-device environments where post-deployment learning and ease of customization are often desirable. 
\end{abstract}


\section{Introduction}
Keyword spotting (KWS) also known as spoken term detection or Keyphrase / Wakeword / Hotword recognition is a task of identifying whether a speech segment contains an utterance of a target keyword. It is typically formulated as a task of classifying short speech segments into one of the pre-defined keywords. Recent advances in neural architectures for on-device, low-footprint KWS models~\cite{sainath2015convolutional,lin2020training,gao2020towards,chen2014small} have made their deployment possible in various commercial products such as Google Home, Apple HomePod, and Amazon Echo. KWS models are typically trained on thousands of hours of in-house datasets \cite{gao2020towards}, often restricted to examples from a very particular set of keywords, limiting their transferability to a larger domain of unseen keywords. However, there has been a rising focus towards on-device personalization of speech recognition technology \cite{sim2019investigation,GoogleAI29:online,personalizedMcGraw}. Enabling keyword spotters to learn from just a few examples would allow easy registration of user-specified keywords on personal devices. \\
\noindent
Towards this goal, we present {\sysname}, a speech embedding model which allows for more accurate KWS models to be learned from fewer training examples. {\sysname} maps short spans of speech utterances to fixed dimensional vectors (embeddings). {\sysname} is  pre-trained on a classification task of recognizing 15K different keywords extracted from the LibriSpeech corpus~\cite{librispeech}. We show that pre-training {\sysname} to simultaneously discriminate across a wide range of 15K keywords for classification allows it to learn more discriminative representations of speech segments. In comparison to recent self-supervised pre-training methods for learning general purpose speech representations~\cite{saeed2020contrastive,shor2020towards}, our supervised pre-training method is directly aligned with the task of keyword spotting. We conduct extensive comparisons between our method and the multi-task supervised pre-training method of \citet{lin2020training}, where each classification head learns to discriminate across just 40 keywords. Despite using three orders of magnitude less data than \citet{lin2020training} during pre-training, we achieve superior performance on several downstream KWS datasets, suggesting that our pre-training method offers better representations by exploiting a much larger vocabulary of keywords. 

Section~\ref{sec:approach} describes the architecture of {\sysname} and the pre-training details. Section~\ref{sec:experiments} describes our experiments on datasets spanning English and four other languages (Japanese, Portuguese, Polish and Esperanto). Compared to recent KWS models, utilizing {\sysname} offers up to 61\% absolute improvements in accuracy, being particularly effective when training data is limited.  Interestingly, even though we pre-train {\sysname} only on English speech, performance gains are also observed on datasets from the four other languages, suggesting that representations learned while pre-training {\sysname} generalize well for the task of keyword spotting.

\section{Related Work}
    {\bf KWS Models}: Keyword spotting is generally formulated as a task of classifying fixed length speech segments into a known vocabulary of keywords. Modern KWS models~\cite{sainath2015convolutional,chen2014small,majumdar2020matchboxnet,bluche2020small,lugosch2018donut,coucke2019efficient,kim2019query,tang2020howl} are often based on neural network classifiers with audio features like Mel-frequency Cepstral Coefficients, extracted from the speech signal as input. \citet{sainath2015convolutional} were among the first to explore CNNs for KWS under memory and compute constrained settings. More recent work by \citet{rybakov2020streaming} provides us with an extensive comparison of a diverse range of KWS architectures. \\
    \noindent
    {\bf KWS with limited data: }Neural network based models require thousands of training examples per keyword to learn useful KWS models. To address this problem, researchers have tried several approaches, including initializing a part of KWS models with weights of an ASR model~\cite{bluche2020predicting}, data-augmentation~\cite{gao2020towards}, meta-learning~\cite{chen2018investigation}, few-shot learning~\cite{parnami2020few} and multitask learning~\cite{lin2020training}. \citet{gao2020towards} propose data augmentation techniques such as the addition of reverberation and noise to simulate far-field speech. SpecAugment~\cite{specaugment2019} is another common data-augmentation technique used in various KWS models~\cite{majumdar2020matchboxnet,rybakov2020streaming}.
    ~\citet{chen2018investigation} extend the few-shot meta-learning framework MAML~\cite{finn2017model} to KWS. ~\citet{parnami2020few} explore prototypical networks~\cite{snell2017prototypical} to learn speech embeddings for improved few-shot transfer. ~\citet{bluche2020predicting} propose a few-shot learning method for adaptation to new keywords. However, they assume access to a grapheme-to-phoneme lexicon, making their approach language-dependent and challenging to extend to low-resource languages where the required lexicons may not be readily available. In contrast, our method does not assume access to such resources and works well even for the languages unseen during the pre-training phase.

    The most similar method to our work is that of~\citet{lin2020training}, where a speech embedding model is learned from 111K hours of speech extracted from YouTube. Their embedding model is pre-trained in a multitask learning setup where it is shared across 125 keyword spotters, each with a different keyword vocabulary of size 40. In contrast, our embedding model is pre-trained on a single task of classifying speech segments into a much larger vocabulary of 15K keywords, thereby learning more discriminative features useful for recognizing keyword utterances. Our embedding model offers significantly better downstream performance across several datasets despite using three orders of magnitude less data during pre-training.
    
\section{Our Method}
\label{sec:approach}
{\bf Pre-training {\sysname}}: We model {\sysname} as neural network $\embeddingFn: U \mapsto {\mathbb R}^d $, that maps the space $U$ of fixed-length speech segments to $d$-dimensional real vectors. $\embeddingFn$ is pre-trained as part of a larger model $P_{\phi}(y | \embeddingFn(u))$ that learns to classify a speech segment $u$ into one of the keywords $y$ belonging to a large vocabulary $\vocab$. Here, $\phi$ represents the parameters of a linear layer before the SoftMax activation that outputs the probability distribution $P_\phi$. Learning to discriminate between many possible keywords simultaneously during classification helps $\embeddingFn$ learn strongly discriminative representations. Training on a diverse range of keywords also allows $\embeddingFn$ to generalize well to the utterances of previously unseen keywords.  We assume that the pre-training data $\trainSet$ is available in the form of keywords paired with their utterances: $\trainSet = \{ (y, U_y) \; | \; y \in \vocab\}$.  $U_y$ represents the example utterances for the keyword $y \in \vocab$. Section~\ref{sec:training_emb} provides details about how we created $\trainSet$. The parameters $\theta$ and $\phi$ are learned jointly by minimizing the cross-entropy loss $\mathcal{L}$ over the entire dataset $\trainSet$ as per Equation~\ref{eqn:pretrain_loss}. 
\begin{equation}
    \label{eqn:pretrain_loss}
    \mathcal{L} = - \sum_{y \in \vocab} \sum_{u \in U_y} \log(P_{\phi}(y | \embeddingFn(u)))
\end{equation}
We minimize the loss using Adam Optimizer \cite{kingma2014adam}, with a learning rate of $5e$$-$$4$  and a batch size of 1024, for 1M steps. We pick the best embedding model as per the dev set described in Section~{\ref{sec:training_emb}} for downstream experiments. \\

\noindent
{\bf Training KWS models}: After pre-training, KWS models are learned by replacing the linear-layer $\phi$ with a randomly initialized linear layer and minimizing the cross-entropy loss on the target KWS dataset. We experiment with both fine-tuning and freezing weights of $\embeddingFn$ during training.\\

\begin{center}
\begin{figure*}[h]
\centering
\subfloat[]{\includegraphics[scale=0.22]{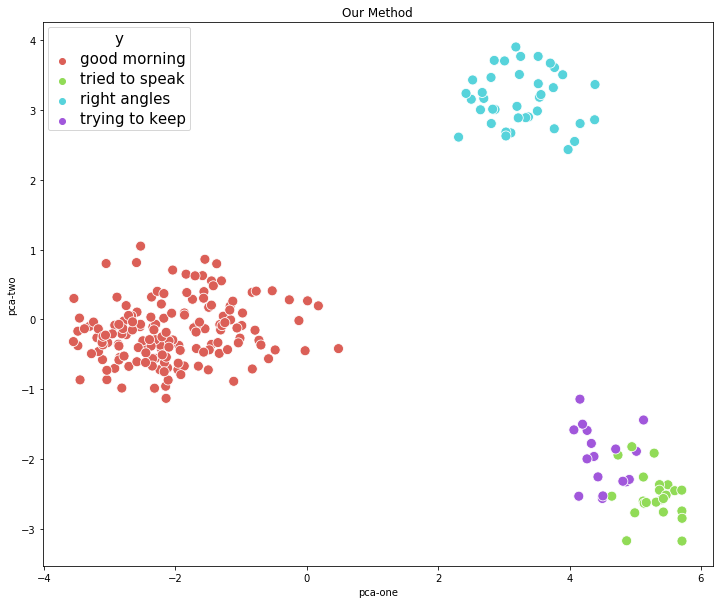}}
\subfloat[]{\includegraphics[scale=0.22]{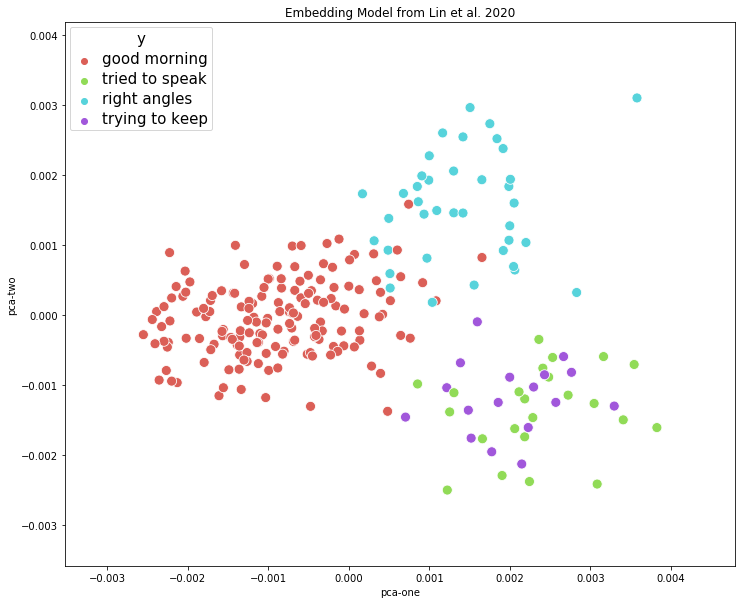}}
\subfloat[]{\includegraphics[scale=0.22]{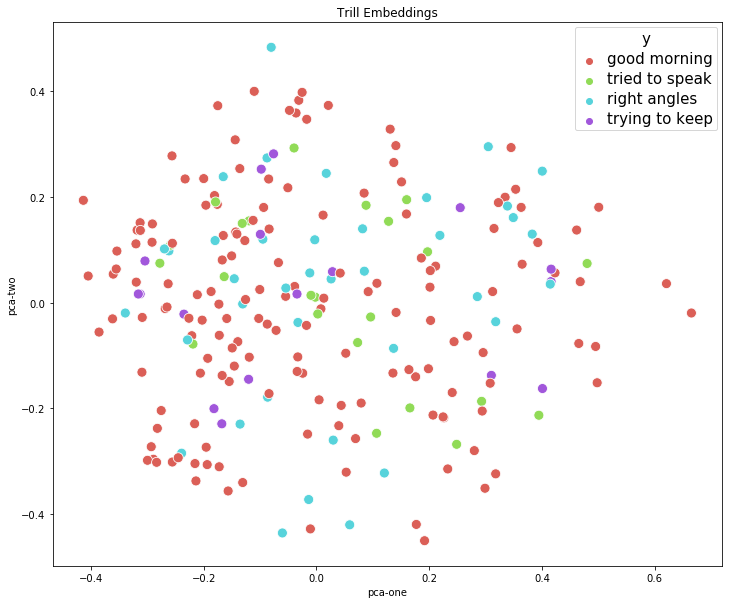}}
\vspace{-1em}
\caption{PCA projections of embeddings obtained from (a) {\sysname}, (b) \citet{lin2020training}, and (c) TRILL~\cite{shor2020towards} for utterances of ``good morning'', ``tried to speak", ``trying to keep", and ``right angles". {\bf Note}: Embedding models were NOT trained on these keywords.
\vspace{-.75em}
}
\label{fig:pca_embeddings}
\end{figure*}
\end{center}

\noindent
Figure~\ref{fig:pca_embeddings} compares the PCA projections of the speech embeddings obtained from (a) {\sysname}, (b) the embedding model in ~\citet{lin2020training}, and (c)  TRILL~\cite{shor2020towards}. TRILL is a speech embedding model pre-trained using an unsupervised triplet-loss objective for learning general purpose speech representations. Observe that {\sysname} provides well-separated clusters for embeddings of different keywords. Even for the similar-sounding keywords like {\tt tried to speak} (in green) and {\tt trying to keep} (in violet), the clusters obtained from {\sysname} are separated reasonably in contrast to the highly overlapping clusters obtained from \citet{lin2020training}'s embedding model. We do not observe any clustering for embeddings obtained from TRILL. Unlike {\sysname}, the loss objective in TRILL is not aligned directly with the task of  discriminating between  keyword utterances. Note that none of the embedding models were pre-trained on utterances of the keywords displayed in Figure~\ref{fig:pca_embeddings}. 
\vspace{-0.5em}
\subsection{Architecture details}
\vspace{-0.25em}
The embedding model $\embeddingFn$ uses a CNN based architecture similar to \cite{lin2020training}. It assumes a 2s long speech segment as input which is converted to 40 dimensional log-mel features obtained using window size of 25 ms and a step size of 10 ms covering the frequency range from 60 Hz to 7800 Hz. The log-mel features are then passed to the CNN architecture which consists of 6 convolutional blocks. The first five convolutional blocks each consist of 5 layers: 4 alternating a 1x3 and 3x1 convolutions, followed by a maxpool layer. After the fifth block the frequency dimension becomes 1. The final block comprises two consecutive layers each composed of a 5x1 convolution and an average-pool layer. Starting with 24 channels in the first block, we increase the number of channels by 24 in each new block until a maximum of 96
is reached. Ultimately, the embedding model maps 2s long speech segments to 96-dimensional feature vectors. We refer the reader to Figure~1 in \cite{lin2020training} for a view of the embedding model's architecture. 

\vspace{-0.5em}
\subsection{Dataset extraction}
\vspace{-0.25em}
\label{sec:training_emb}
Pre-training {\sysname} requires keyword utterances from a large representative vocabulary $\vocab$. The publicly available LibriSpeech corpus~\cite{librispeech}, together with community created forced-alignments~\cite{alignmentsLibrispeech,mcauliffe2017montreal}, provide us with many easily extractable keywords. As keywords, we use $n$-grams ($n<=5$) that are at least 10 characters long and have at least 10 occurrences in the train split of LibriSpeech corpus, giving us over 15.2K different keywords. The extracted keyword utterances constitute 250hrs of speech. Keyword utterances in the dev split of LibriSpeech serve as the dev set during pre-training.

Prior datasets used for benchmarking KWS models' performance involve only a small number of keywords (often less than 20). To evaluate KWS models on a broader range of unseen keywords, we hold out 150 keywords with at least 100 occurrences and their associated utterances from the pre-training dataset. 
In Section~\ref{sec:experiments}, we report the performance of various models on this held out dataset, which we refer to as $\librikws$.
\vspace{-2.0em}

\section{Experiments}
\label{sec:experiments}
Our experiments on three datasets spanning five languages show that {\sysname} helps tremendously when training data is limited, even for the languages not seen by {\sysname} during pre-training. This section provides a brief description of the datasets followed by the experimental results. We plan to release the datasets and pre-trained models.
\vspace{-0.5em}
\subsection{Datasets}
\label{sec:datasets}
\vspace{-0.25em}
{\bf Google Speech Commands (V2)} \cite{warden2018speech} is  
commonly used for benchmarking KWS models. It consists of utterances for 35 different words like {\{\tt Up, Down, Left, Right, ...\}}. 25 of these words are grouped into a single category representing the {\tt negative} class. 
We use the default train, dev, and test splits as available on the tensorflow website~\cite{speechcommands_web} consisting of 85.5K, 10.1K and 4.9K utterances respectively. \\
{\bf $\librikws$:}  
As described in Section~\ref{sec:training_emb}, {\librikws} contains utterances for the 150 keywords not used to pre-train {\sysname}. We use 90 examples per keyword for the train split and 10 examples per keyword for the dev split. For evaluation, we construct a `test-clean' set by pooling utterances extracted from the `test-clean' and `dev-clean' splits of LibriSpeech. Similarly, a `test-other' set is constructed by pooling utterances extracted from the `test-other' and `dev-other' splits. The `test-other' split is expected to be more challenging than `test-clean'. The design of the original LibriSpeech dataset~\cite{librispeech} ensures zero speaker overlap across between the train and test splits. To the best of our knowledge, no open-source KWS dataset provides more target classes than $\librikws$. \\
{\bf Common Voice}~\cite{commonvoice}: To evaluate the effectiveness of {\sysname} on non-English keywords, we use the single word spoken digit recognition dataset for four languages, including Japanese (ja), Esperanto (eo), Polish (pl), and Portuguese (pt). The number of target keywords varies between 12 and 14. We use the default train, test, and dev splits consisting an average of 335, 311, and 306 utterances respectively. 
 
To facilitate comparisons with prior work and given no severe class imbalances in the test sets, we report accuracy as a performance metric. 

\subsection{Results}
\label{sec:results}
{\bf Overall Comparisons: } Table~\ref{tab:overall_results} presents our method's overall comparisons with recent KWS models after training each model on the entire training set of the respective datasets. Our comparisons include MatchBoxNet~\cite{majumdar2020matchboxnet} composed of convolutional layers with residual connections, with roughly 85K weights. MHAtt-RNN represents the most accurate model from~\cite{rybakov2020streaming} with roughly 765K weights, composed of multi-head attention layers~\cite{vaswani2017attention} followed by an RNN layer. For MatchBoxNet and MHAtt-RNN, we use the implementation provided by~\cite{rybakov2020streaming,tfmatchbox}. MTLEmb refers to KWS models using convolutional heads on top of the speech embedding model described in~\cite{lin2020training}, with roughly 400K weights in the embedding model and 50K weights in the convolutional head. {\sysname} has roughly 410K weights and requires just 96 additional weights per keyword in the linear classification layer. We experiment with fixing (fix) and finetuning (FT) the weights of the embedding models during training. {\sysname} (rand) refers to the KWS model obtained by training a randomly initialized {\sysname} architecture from scratch. We observe that utilizing a pre-trained {\sysname} provides the best results across all the datasets. Keeping {\sysname}'s weights fixed while training KWS models also yields competitive results and requires only 96 trainable parameters per keyword. Since the train-split of {\librikws} contains only 90 examples per keyword, fixing {\sysname}'s weights also helps avoid over-fitting and yields superior results than finetuning the entire model.
%\vspace{-0.5em}
\begin{table}[bt]
\caption{Comparison of various KWS models on the Speech Commands test set (SC), test-clean (LK-c) and test-other (LK-o) splits of LibriKWS, and the test splits of four selected languages in the Common Voice dataset. Models are trained using all the data in respective train splits and the accuracy measured.}
\vspace{-0.5em}
\begin{adjustbox}{width=\columnwidth,center}
\begin{tabular}{|l|r|r|r|r|r|r|r|}
\hline
Method         & SC   & LK-c & LK-o &  ja   & eo   & pl   & pt   \\ \hline \hline
Matchbox       & 98.0 & 97.3  & 89.8   & 76.0 & 88.5 & 85.0 & 84.3 \\ \hline
MHAtt-RNN      & 98.0 & 99.7  & 95.3   & 86.0 & 87.0 & 87.3 & 79.3 \\ \hline
MTLEmb (fix)  & 96.6 & 95.1  & 87.2   & 86.7 & 87.4 & 89.5 & 82.6 \\ \hline
MTLEmb (FT)   & 97.7 & 94.9  & 88.1   & 75.0 & 79.6 & 74.2 & 72.1 \\ \hline
{\sysname} (rand) & 97.2 & 93.6  & 78.1   & 83.2 & 84.1 & 85.4 & 78.3 \\ \hline
{\sysname} (fix)     & 93.9 & \textbf{99.8}  & \textbf{97.8}   & 92.3 & \textbf{91.2} & \textbf{90.3} & 82.4 \\ \hline
{\sysname} (FT)      & \textbf{98.2} & 97.8  & 93.2   & \textbf{92.9} & 89.1 & \textbf{90.3} & \textbf{84.7} \\ \hline
\end{tabular}
\end{adjustbox}
%\vspace{-1.0em}
\label{tab:overall_results}
\vspace{-1.25em}
\end{table}

\\

\begin{figure}[t]
    %\vspace{-1.0em}
    \includegraphics[scale=0.23]{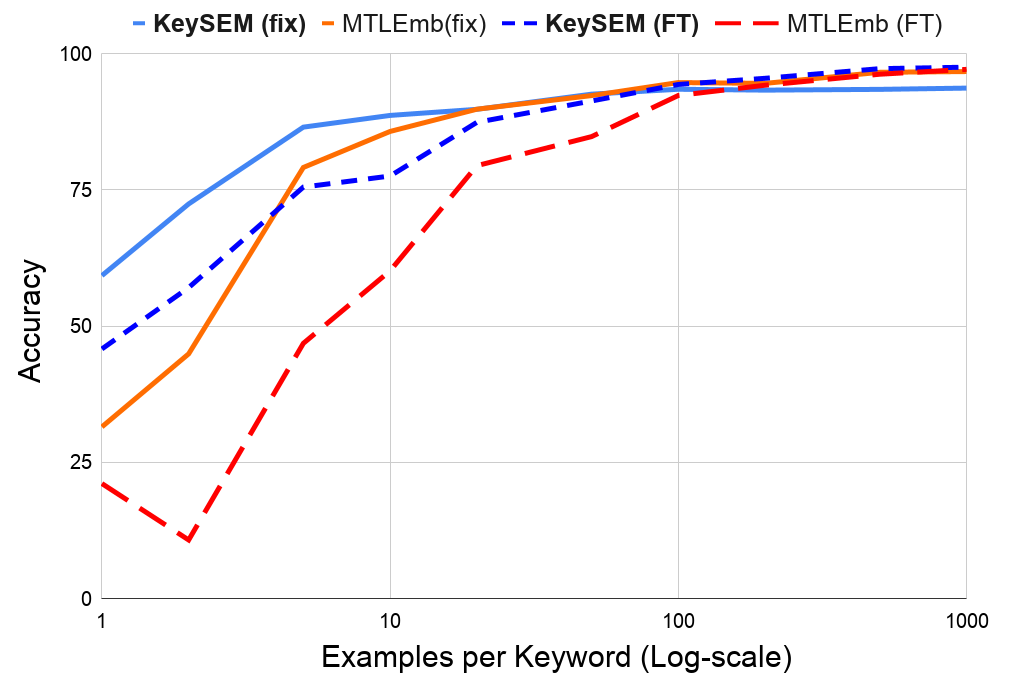}
    %\vspace{-1.0em}
    \caption{Performance comparison between {\sysname} and {\linetalmp} based KWS models on Speech Commands dataset with varying number of trianing examples per keyword.}
    \label{fig:sc_ablation}
    \vspace{-2.25em}
\end{figure}

%\vspace{-1.0em}
\noindent
{\bf Limited data experiments:} Table~\ref{tab:overall_5} reports the performance of various KWS models when trained on just 5 randomly selected examples per keyword. For the Speech Commands dataset, we retain 15 training examples for the negative class. Models like Matchbox and MHAtt-RNN utilize SpecAugment~\cite{specaugment2019} for data augmentation to overcome data scarcity and overfitting. From Table~\ref{tab:overall_5}, we observe that utilizing pre-trained embedding models like~\citet{lin2020training} or {\sysname} with fixed weights provides consistent gains over fine-tuning or training the models from scratch using data augmentation techniques. {\sysname} offers absolute gains between 7\% to 61\% over \citet{lin2020training}'s method (\linetalmp). Surprisingly, the gains offered by {\sysname} also extend well to datasets from four other languages, including Japanese (ja), Esperanto (eo), Polish (pl), and Portuguese (pt), even though {\sysname} is pre-trained only on English utterances. Figure~\ref{fig:sc_ablation} provides a closer comparison between {\sysname} and {\linetalmp}. Along the x-axis, we increase the number of training examples per keyword and report the test accuracy on the y-axis for the Speech Commands dataset. With less than 10 training examples per keyword, training with {\sysname}'s weights held fixed offers superior performance over other methods. 
The performance difference across various methods diminishes with increasing amounts of training data. For 1000 examples per keyword, fine-tuning the entire {\sysname} offers the highest accuracy of 97.5\%.\\

\noindent
Although \citet{lin2020training} utilize three orders of magnitude more data during pre-training, the two main reasons behind superior performance of {\sysname} are as follows. (i)~Learning to simultaneously discriminate across 15K keywords during pre-training allows {\sysname} to develop more discriminative features in comparison to the pre-training method of \cite{lin2020training} where individual classifiers learn to discriminate between just 40 keywords. (ii)~{\sysname} is pre-trained to perform classification with a linear layer that requires only 96 trainable parameters per keyword. In contrast, the embedding model in \cite{lin2020training} is pre-trained to perform classification with convolutional heads containing 50K parameters each. Even if the embedding model is held fixed, the convolutional heads are more susceptible to over-fitting than linear layers when training data is limited. Training a linear layer with the embedding model in \cite{lin2020training} leads to poor accuracy of less than 70\% on Speech Commands even after utilizing all the training examples, which indicates that their embedding model is pre-trained to work best with convolutional heads. 

\begin{table}[t]
\caption{Comparing accuracies of various KWS models on the Speech Commands test set (SC), test-clean (LK-c) and test-other (LK-o) splits of LibriKWS, and the test splits of four selected languages in the Common Voice dataset. \textbf{Note}: Models were trained using only 5 randomly chosen examples per keyword from the respective train sets.}
\vspace{-0.5em}
\begin{adjustbox}{width=\columnwidth,center}
\begin{tabular}{|l|r|r|r|r|r|r|r|}
\hline 
Method         & SC   & LK-c & LK-o &  ja   & eo   & pl   & pt   \\ \hline \hline
Matchbox       & 45.3 & 3.0  & 2.8   &  48.0 &	60.5 &	58.3 &	36.3  \\ \hline
MHAtt-RNN      & 48.2 & 27.3  &  12.8  &  55.0 &	62.5  &	68.0 &	53.0  \\ \hline
MTLEmb (fix)  & 79.1  & 43.9  &  33.1  &  76.5 & 74.5 &	76.3 &	71.3  \\ \hline
MTLEmb (FT)   & 46.8  & 30.1   & 21.4  &  41.3 &	48.0  &	43.8 &	37.6  \\ \hline
{\sysname} (fix) &  \textbf{86.5}  & \textbf{98.5}  & \textbf{94.5}  & \textbf{86.2} & \textbf{87.4} & \textbf{86.3} & \textbf{80.8} \\ \hline
{\sysname} (FT) &  75.5  & 62.3  &  44.4  & 82.1 &	82.7 &	83.9 &	75.8 \\ \hline
\end{tabular}
\end{adjustbox}
%\vspace{-1.0em}
\label{tab:overall_5}
%\vspace{-2.0em}
\end{table}
\vspace{-0.5em}
\subsection{Learning new keywords sequentially}
\begin{figure}[t]
    %\vspace{-0.5em}
    \centering
    \includegraphics[scale=0.30]{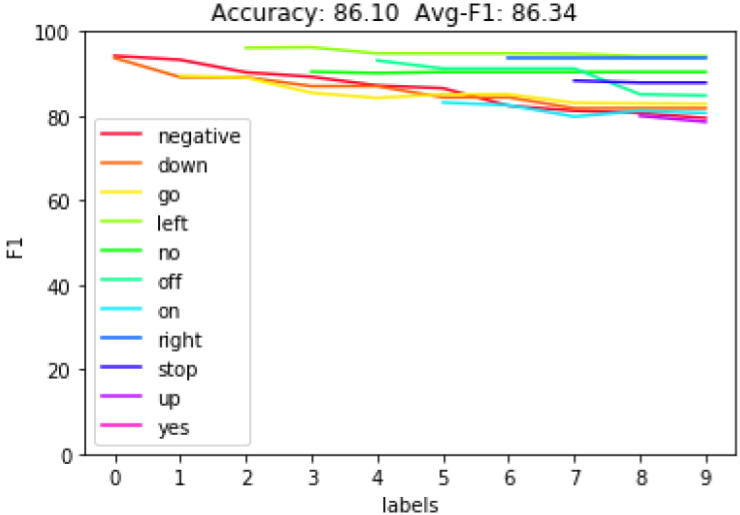}
    \vspace{-0.5em}
    \caption{Learning new keywords sequentially}
    \label{fig:incremental_kws}
    \vspace{-2.0em}
\end{figure}

\vspace{-0.25em}
Learning new keywords over time, with just a few examples, is desirable for customizing KWS models as per user needs. We have observed that simply training linear classifiers with {\sysname}'s weights held fixed provides superior performance in low-data conditions. Not requiring to finetune {\sysname} makes it particularly attractive when examples for new keywords arrive sequentially. Finetuning neural networks for learning new classes sequentially has been shown to cause catastrophic forgetting of previously learned classes~\cite{rebuffi2017icarl,kirkpatrick2017overcoming}. In this section, we use the Speech Commands dataset to explore the ability of {\sysname} to learn new keywords sequentially. We begin with learning a sigmoid classifier to distinguish between the utterances of the {\tt negative} and {\tt down} label. Examples for subsequent classes arrive in sequential order, as shown in the legend of Figure~\ref{fig:incremental_kws}. To account for memory constraints in practical use-cases, we only assume access to examples of the {\tt negative} class and the newly observed class while training. Access to examples of previously learned classes is not assumed, which rules out the possibility of re-training the classifier using all the observed data. We learn an independent sigmoid classifier for each new class, which learns to distinguish the new class from the {\tt negative} class. For training, we assume only 50 examples for the {\tt negative} class and only 5 examples each for other keywords. During inference when predicted confidence for all the keywords is less than 0.5, we predict output as {\tt negative}, otherwise we output the keyword with maximum confidence. After learning all the 10 classes, the overall accuracy of the model is 86.1\%, only 0.4\% worse than training on examples of multiple classes simultaneously (Table~2, {\sysname}(fix)). Figure~\ref{fig:incremental_kws} shows variation of the F1 score of previously learned classes as data for new classes arrives along the x-axis. We observe a gradual drop in performance of previous classes while learning a new class. However, the drop is not catastrophic. In contrast, if weights of the embedding model are finetuned, the accuracy of previous classes drop catastrophically while new classes are learned, the overall final accuracy being just 24.3\%.\\
Post-deployment learning of new keywords with limited examples is often desirable in an on-device set-up. {\sysname} enables learning reasonably accurate KWS models under such conditions, without relying on complex methods like~\cite{rebuffi2017icarl,kirkpatrick2017overcoming} to avoid catastrophic forgetting. We defer exploration of methods like~\cite{rebuffi2017icarl,kirkpatrick2017overcoming}  in the context of KWS to future work.  

\vspace{-0.25em}
\section{Conclusion}
\vspace{-0.25em}
We propose {\sysname}, a speech embedding model pre-trained to recognize utterances of a large number of keywords. Learning to discriminate across a wide range of keywords during pre-training allows {\sysname} to offer features helpful for the task of Keyword Spotting. On datasets spanning five different languages, {\sysname} offers consistent performance gains over several recent methods.
When training data is limited, {\sysname} achieves superior performance over other methods by training only a linear classifier's parameters. Finally, considering post-deployment learning in on-device environments, we demonstrate {\sysname}'s effectiveness in learning from a stream of data where examples for new classes arrive sequentially.

\noindent
{\bf Acknowledgement} The authors would like to thank the anonymous reviewers,  Andre Perunicic, Dominik Roblek, Matt Sharifi, Marco Tagliasacchi, Karolis Misiunas, Mauro Verzetti, and Vihari Piratla, for their constructive feedback on this work. 

\bibliographystyle{IEEEtranN}

\bibliography{mybib}

\end{document}